\documentclass[11pt]{article}

\usepackage{amsmath}
\usepackage{graphicx}
\usepackage{amsfonts}
\usepackage{amssymb}
\usepackage{epsfig}
\usepackage{color}
\usepackage{psfrag}
\usepackage{epstopdf}

\newcommand{\EQ}{\begin{equation}}
\newcommand{\EN}{\end{equation}}
\newcommand{\be}{\begin{equation}}
\newcommand{\ee}{\end{equation}}
\newcommand{\bea}{\begin{eqnarray}}
\newcommand{\eea}{\end{eqnarray}}

\begin{document} \setcounter{page}{0}
\topmargin 0pt
\oddsidemargin 5mm
\renewcommand{\thefootnote}{\arabic{footnote}}
\newpage
\setcounter{page}{0}
\topmargin 0pt
\oddsidemargin 5mm
\renewcommand{\thefootnote}{\arabic{footnote}}
\newpage
\begin{titlepage}
\begin{flushright}
SISSA 07/2013/FISI \\
\end{flushright}
\vspace{0.5cm}
\begin{center}
{\large {\bf Interfaces and wetting transition on the half plane.}\\
{\bf Exact results from field theory}}\\
\vspace{1.8cm}
{\large Gesualdo Delfino and Alessio Squarcini}\\
\vspace{0.5cm}
{\em SISSA -- Via Bonomea 265, 34136 Trieste, Italy}\\
{\em INFN sezione di Trieste}\\
\end{center}
\vspace{1.2cm}

\renewcommand{\thefootnote}{\arabic{footnote}}
\setcounter{footnote}{0}

\begin{abstract}
\noindent
We consider the scaling limit of a generic ferromagnetic system with a continuous phase transition, on the half plane with boundary conditions leading to the equilibrium of two different phases below criticality. We use general properties of low energy two-dimensional field theory to determine exact asymptotics of the magnetization profile perperdicularly to the boundary, to show the presence of an interface with endpoints pinned to the boundary, and to determine its passage probability. The midpoint average distance of the interface from the boundary grows as the square root of the distance between the endpoints, unless the reflection amplitude of the bulk excitations on the boundary possesses a stable bound state pole. The contact angle of the phenomenological wetting theory is exactly related to the location of this pole. Results available from the lattice solution of the Ising model are recovered as a particular case.
\end{abstract}
\end{titlepage}

\newpage
\section{Introduction}
Interfacial phenomena at boundaries are a subject of relevant interest for both theory and applications. On the theoretical side, the one this paper is concerned with, the effects of the boundary on an interface separating different phases of a statistical system have been extensively studied using phenomenological, mean field, renormalization group and other approximation methods ([1-8] is a certainly incomplete list of review articles). The only exact result that has been available concerns the Ising model on the half plane \cite{Abraham_wetting,Abraham}, a circumstance that, while confirming a specificity of the two-dimensional case, raises the question about the role of Ising solvability in these exact findings. 

We show in this paper that exact results including those of \cite{Abraham} as a particular case are obtained quite generally for any two-dimensional model exhibiting a continuous phase transition. This is done extending to the half plane the non-perturbative field theoretical approach recently used in \cite{DV} to study phase separation on the whole plane. As in that case, general exact results emerge because, when its end-to-end distance $R$ is much larger than the correlation length, the interface is described by a single particle (domain wall) state, in a low energy limit leading to a general solution. In this way, the fluctuations of the interface turn out to be ruled by the low energy singularity of the matrix element of the order parameter field (as for the whole plane), with the fields pinning the interface endpoints to the boundary producing boundary reflection and an average midpoint distance from the boundary of order $\sqrt{R}$. 

The result changes qualitatively if boundary and domain wall excitation admit a stable bound state, which becomes dominant in the spectral sum at low energies and bounds the interface to the boundary. The contact angle and the spreading coefficient of the phenomenological theory of wetting then emerge in a completely natural way within the field theoretical formalism.

The paper is organized as follows. In the next section we illustrate the 
field theoretical setting and derive the results for the unbound interface.
Section~3 is then devoted to the effects produced by the bound state and to the characterization of the wetting transition, while section~4 contains some final remarks.

\section{Interfaces on the half plane}
Consider a ferromagnetic spin model of two-dimensional classical statistical mechanics in which spins take discrete values labelled by an index $a=1,2,\ldots,n$. The energy of the system is invariant under global
transformations of the spins according to a symmetry whose spontaneous breaking below a critical temperature $T_c$ is responsible for the presence on the infinite plane of $n$ translation invariant pure phases; we denote $\langle\cdots\rangle_{a}$ statistical averages in the phase $a$. 

Assuming a continuous transition, we consider the scaling limit below $T_c$, corresponding to a Euclidean
field theory defined on the plane with coordinates $(x,y)$, which can be seen as the analytic continuation to imaginary time of a (1+1)-dimensional relativistic field theory with space coordinate $x$ and time coordinate $t=iy$. If $H$ and $P$ are the Hamiltonian and momentum operators and $\Phi$ a field of the theory, translation invariance on the plane yields the relation
\EQ
\Phi(x,y)=e^{ixP+yH}\Phi(0,0)e^{-ixP-yH}\,.
\label{translations}
\EN
The (1+1)-dimensional theory possesses degenerate vacua $|0\rangle_a$ associated to the pure phases of the system. The elementary excitations correspond to stable kink states $|K_{ab}(\theta)\rangle$ interpolating between different vacua
$|0\rangle_a$ and $|0\rangle_b$. We introduced the rapidity variable $\theta$ which conveniently parameterizes the energy and momentum of the kinks as $(E,p)=(m\cosh\theta,m\sinh\theta)$, $m$ being the kink mass or inverse correlation length. The trajectory of the kink on the Euclidean plane corresponds to a domain wall between the phases $a$ and $b$. Multi-kink excitations take the form $|K_{aa_1}(\theta_1)K_{a_1a_2}(\theta_2)\ldots K_{a_{n-1}b}(\theta_n)\rangle$. Within the scattering framework \cite{ELOP} we consider, these are asymptotic states, incoming if considered long before the collisions among the kinks, outgoing if considered long after, and their energy is simply $\sum_{i=1}^nm\cosh\theta_i$.

\begin{figure}[t]
\begin{center}
\includegraphics[width=9cm]{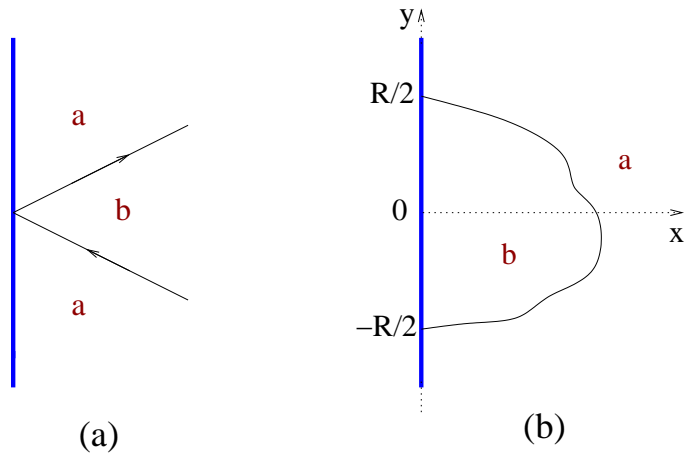}
\caption{Elastic scattering (reflection) of a kink off the boundary (a), and interface pinned at the boundary (b).}
\label{wet_merge}
\end{center}
\end{figure}

Consider now the system on the half-plane $x\geq 0$. We denote by $B_a$ a boundary condition at $x=0$ which is $y$-independent and breaks the symmetry of the bulk in the direction $a$ in order parameter space; this can be realized applying a constant boundary magnetic field pointing in the direction $a$. We denote $\langle\cdots\rangle_{B_a}$ statistical averages in presence of the boundary condition $B_a$. Preservation of translation invariance in the $y$ direction yields energy conservation in the $(1+1)$-dimensional picture. 
The bulk excitations are still the kink states described for the full plane case, but now they are restricted to $x>0$; we indicate this restriction by a subscript $B_a$. Hence $|0\rangle_{B_a}$ denotes the vacuum (no excitations in the bulk) on the half-plane with the boundary condition $B_a$. If $\sigma$ is the spin field, the magnetization $\langle\sigma(x,y)\rangle_{B_a}={}_{B_a}\langle 0|\sigma(x,y)|0\rangle_{B_a}$ points in the direction $a$ and depends only on the distance $x$ from the boundary; in particular
\EQ
\lim_{x\to\infty}\langle\sigma(x,y)\rangle_{B_a}=\langle\sigma\rangle_a\,,
\label{bulk_magnetization}
\EN
where $\langle\sigma\rangle_a$ is the constant magnetization in phase $a$ on the full plane. The state $|0\rangle_{B_a}$ is eigenstate of the Hamiltonian $H_{B_a}$ of the system on the half line. We consider the case in which boundary conditions $B_a$ and $B_b$ are related by the symmetry, so that $|0\rangle_{B_a}$ and $|0\rangle_{B_b}$ have the same energy $E_B$.

The asymptotic scattering state $|K_{ba}(\theta)\rangle_{B_a}$ corresponds to an incoming kink (travelling towards the boundary) if its momentum is negative, i.e. if $\theta<0$. If its energy is lower than the energy $2m$ needed to produce two kinks upon interaction with the boundary, it will simply be reflected into an outgoing kink\footnote{As emphasized in \cite{GZ}, the analogies between bulk and boundary scattering become evident thinking of the boundary as the propagation of an infinitely heavy particle sitting at $x=0$.} with rapidity $-\theta$ (Fig.~1a). The state $|K_{ba}(\theta)\rangle_{B_a}$ is eigenstate of $H_{B_a}$ with eigenvalue $E_{B}+m\cosh\theta$.

We are now ready to set up the configuration we want to study, namely a boundary condition which is of type $B_a$ if $|y|>R/2$ and of type $B_b$ if $|y|<R/2$. The interest of such a boundary condition, that we denote $B_{aba}$, is easily understood observing that the limit for $x\to\infty$ of the magnetization profile $\langle\sigma(x,0)\rangle_{B_{aba}}$ has to tend to $\langle\sigma\rangle_a$ if $R$ is finite, and to $\langle\sigma\rangle_b$ if $R$ is infinite. The natural way to account for this situation is to expect the formation of an interface pinned at $R/2$ and $-R/2$ on the boundary, separating an inner phase $b$ from an outer phase $a$ (Fig.~1b), and whose average distance from the boundary at $y=0$ diverges with $R$. The remainder of this section is devoted to see how such a picture indeed emerges within our general field theoretical framework.


Technically the change from the boundary condition $B_a$ to $B_b$ at a point $y$ is realized starting with $B_a$ and inserting on the boundary a field $\mu_{ab}(0,y)$ which acting on the vacuum $|0\rangle_{B_a}$ creates kink states interpolating between phase $a$ and phase $b$. Hence the simplest non-vanishing matrix element of the boundary field $\mu_{ab}$ is
\EQ
{}_{B_a}\langle 0|\mu_{ab}(0,y)|K_{ba}(\theta)\rangle_{B_a}=e^{-ym\cosh\theta}
{}_{B_a}\langle 0|\mu_{ab}(0,0)|K_{ba}(\theta)\rangle_{B_a}\equiv e^{-ym\cosh\theta}{\cal F}_\mu(\theta)\,.
\label{bff}
\EN
The partition function of the system with boundary condition $B_{aba}$ reads
\EQ
Z={}_{B_a}\langle 0|\mu_{ab}(0,R/2)\mu_{ba}(0,-R/2)|0\rangle_{B_a}=\int_0^\infty\frac{d\theta}{2\pi}|{\cal F}_\mu(\theta)|^2 e^{-mR\cosh\theta}+O(e^{-2mR})\,,
\EN
where the last expression is obtained expanding over an intermediate set of outgoing kink states and retaining only the lightest (single kink) contribution which is leading in the large $mR$ limit we will consider from now on. Since the above integral is dominated by small rapidities and ${\cal F}_\mu$ is expected to behave as\footnote{Linear behavior of matrix elements at small rapidities in two-dimensional theories is well known. Within the framework of integrable boundary field theory \cite{GZ} exact examples can be found in \cite{BPT}. More generally, see \cite{Smirnov} about matrix elements in integrable theories.}
\EQ
{\cal F}_\mu(\theta)=a\,\theta+O(\theta^2)\,,
\label{fmu}
\EN
the partition function becomes
\EQ
Z\sim |a|^2\int_0^\infty\frac{d\theta}{2\pi}\,\theta^2\,e^{-mR(1+\theta^2/2)}=\frac{|a|^2\,e^{-mR}}{2\sqrt{2\pi}\,(mR)^{3/2}}\,.
\label{Z} 
\EN
The magnetization profile along the $x$ axis is given by
\bea
&& \langle\sigma(x,0)\rangle_{B_{aba}}=\frac{1}{Z}\,{}_{B_a}\langle 0|\mu_{ab}(0,R/2)\sigma(x,0)\mu_{ba}(0,-R/2)|0\rangle_{B_a}
\label{profile1}\\
&& \sim \frac{1}{Z}\int_{-\infty}^{+\infty}\frac{d\theta_1}{2\pi}\frac{d\theta_2}{2\pi}{\cal F}_\mu(\theta_1)\langle
K_{ab}(\theta_1)|\sigma(0,0)|K_{ba}(\theta_2)\rangle{\cal F}_\mu^*(\theta_2)
e^{m[i(\sinh\theta_1-\sinh\theta_2)x-(\cosh\theta_1+\cosh\theta_2)\frac{R}{2}]}\,,\nonumber
\eea
where in the last line we have taken $mR\gg 1$ to project on the one-kink intermediate states, but also $mx\gg 1$ to be able to treat $\sigma(x,0)$ as a bulk field which satisfies (\ref{translations}) and is evaluated on bulk kink states (whose rapidities take both positive and negative values). In other words, for $mx$ large the only effect of the boundary on the magnetization comes from the boundary changing fields at $(0,\pm R/2)$; in their absence one would simply observe the constant value $\langle\sigma\rangle_a$.
The bulk matrix element of the spin field between one-kink states is related by the crossing relation\footnote{Crossing a particle from the initial to the final state (or vice versa) involves reversing the sign of its energy and momentum \cite{ELOP}, namely an $i\pi$ rapidity shift. The delta function term in (\ref{crossing}) is a disconnected part arising from annihilation of the two kinks.} 
\begin{equation} 
\langle K_{ab}(\theta_1)|\sigma(0,0)|K_{ba}(\theta_2)\rangle={ F}_{\sigma}(\theta_1+i\pi-\theta_2)+2\pi\delta(\theta_1-\theta_2)\langle\sigma\rangle_{a}\,,
\label{crossing}
\end{equation}
to the form factor
\EQ
{F}_{\sigma}(\theta_1-\theta_2)\equiv{}_a\langle 0|\sigma(0,0)|K_{ab}(\theta_1)K_{ba}(\theta_2)\rangle\,.
\label{ff}
\EN
As already observed in \cite{DV} for the case of phase separation on the whole plane, it is crucial that quite generally, due to non-locality of the kinks with respect to the spin field, ${F}_{\sigma}(\theta)$ possesses an annihilation pole at $\theta=i\pi$ with residue \cite{DC98}
\begin{equation}
-i\,\text{Res}_{\theta=i\pi}{F}_{\sigma}(\theta)=\langle\sigma\rangle_{a}-\langle\sigma\rangle_{b}\equiv\Delta\langle\sigma\rangle\,.
\label{residue}
\end{equation}
Since $mR$ is large (\ref{profile1}) is dominated by small rapidities and (\ref{fmu}), (\ref{crossing}) and (\ref{residue}) lead to
\EQ
\langle\sigma(x,0)\rangle_{B_{aba}}\sim 2\langle\sigma\rangle_{a}+i\,\Delta\langle\sigma\rangle\
\frac{|a|^2}{Z}e^{-mR}\int_{-\infty}^{+\infty}\frac{d\theta_1}{2\pi}\frac{d\theta_2}{2\pi}\,\frac{\theta_1\theta_2}{\theta_1-\theta_2}
e^{m[i(\theta_1-\theta_2)x-(\theta_1^2+\theta_2^2)\frac{R}{4}]}\,.
\label{profile2}
\EN
Differentiation removes the singularity of the integrand and gives
\bea
\partial_{mx}\langle\sigma(x,0)\rangle_{B_{aba}}&\sim & -\Delta\langle\sigma\rangle\,\frac{|a|^2 e^{-mR}}{(2\pi)^2 Z}\,g(x)g(-x)\nonumber\\
&=& \Delta\langle\sigma\rangle\,\frac{4\sqrt{2}}{\sqrt{\pi\,mR}}\,z^2\,e^{-z^2}\,,\hspace{2cm}z\equiv\sqrt{\frac{2m}{R}}\,x
\label{derivative}
\eea
where we used (\ref{Z}) and
\EQ
g(x)=\int_{-\infty}^{+\infty}d\theta\,\theta\,e^{-mR\theta^2/4+imx\theta}=\frac{2i\sqrt{2\pi}}{mR}\,z\,e^{-z^2/2}\,.
\EN
Integrating (\ref{derivative}) with the asymptotic condition $\langle\sigma(\infty,0)\rangle_{B_{aba}}=\langle\sigma\rangle_{a}$ gives
\EQ
\langle\sigma(x,0)\rangle_{B_{aba}}\sim \langle\sigma\rangle_{b}-\frac{2}{\sqrt{\pi}}\,\Delta\langle\sigma\rangle\left(z\,e^{-z^2}-\int_0^z du\,e^{-u^2}\right),\hspace{1cm}mx\gg 1\,.
\label{profile}
\EN
From this result we can compute exactly $\lim_{R\to\infty}\langle\sigma(\frac{\alpha}{m}\,(mR)^\delta,0)\rangle_{B_{aba}}$, obtaining $\langle\sigma\rangle_b$ for $0<\delta<1/2$, $\langle\sigma\rangle_a$ for $\delta>1/2$, and the r.h.s. of (\ref{profile}) with $z=\alpha\sqrt{2}$ for $\delta=1/2$. For $\langle\sigma\rangle_a=-\langle\sigma\rangle_b=\langle\sigma\rangle_+$ these are precisely the limits obtained from the lattice in \cite{AI,Abraham_wetting} for the Ising model on the half plane with boundary spins fixed to be positive for $|y|>R/2$ and negative for $|y|<R/2$.

The derivative (\ref{derivative}) of the magnetization profile is peaked around $z=1$, confirming the presence of an interface whose average distance from the boundary increases as $\sqrt{R/m}$. It is also easy to see that the result for the magnetization profile is consistent with a simple probabilistic interpretation. Since we are computing the magnetization on a scale $R$ much larger than the correlation length and far away from the boundary, we can think of the interface as a sharp separation between pure phases\footnote{It has been shown in \cite{DV} how the internal structure of the interface arises from subleading terms in the large $mR$ expansion.}, and write 
\EQ
\langle\sigma(x,0)\rangle_{B_{aba}}\sim\langle\sigma\rangle_a\int^{x}_{0}du~p(u)+\langle\sigma\rangle_b\int_{x}^{\infty}du~p(u),\hspace{1cm}mx\gg 1\,,
\EN
where $p(u)du$ is the probability that the interface intersects the $x$-axis in the interval $(u,u+du)$, so that the two integrals are the left and right passage probabilities with respect to $x$. Differentiating and comparing with (\ref{derivative}) gives the passage probability density
\EQ
p(x)=4\sqrt{\frac{2m}{\pi R}}\,z^2\,e^{-z^2}\,,
\label{probability}
\EN
which correctly satisfies $\int_{0}^{\infty}dx~p(x)=1$.

\section{Wetting transition}
The results of the previous section are modified if the kink-boundary system associated to the asymptotic state $|K_{ab}(\theta)\rangle_{B_b}$ admits a stable bound state $|0\rangle_{B'_{a}}$, corresponding to the binding of the kink $K_{ab}$ on the boundary ${B_{b}}$. As usual for stable bound states \cite{ELOP}, such a binding will correspond to a ``virtual'' value $\theta_0$ of the kink rapidity leading to a bound state energy $E_{B}+m\cosh\theta_0$ real and smaller than the unbinding energy $E_{B}+m$. This amounts to taking $\theta_0=iu$ with $0<u<\pi$, so that 
\EQ
E_{B'}=E_{B}+m\cos u\,.
\label{bs}
\EN
The existence of the bound state manifests in particular through a simple pole in the elastic scattering amplitude of the kink off the boundary, which reads ${\cal R}(\theta)\sim ig^2/(\theta-iu)$ for $\theta\to iu$, with $g$ a kink-boundary coupling constant (Fig.~2a). This pole is inherited by the matrix element (\ref{bff}), for which we have\footnote{Exact solutions exhibiting boundary bound states poles can be found in \cite{GZ} for scattering amplitudes and in \cite{BPT} for matrix elements.} (Fig.~2b)
\EQ
{\cal F}_\mu(\theta)={}_{B_a}\langle 0|\mu_{ab}(0,0)|K_{ba}(\theta)\rangle_{B_a}\sim\frac{ig}{\theta-iu}\,{}_{B_a}\langle 0|\mu_{ab}(0,0)|0\rangle_{B'_{a}}\,,\hspace{1cm}\theta\to iu\,.
\label{ff_pole}
\EN

\begin{figure}[t]
\begin{center}
\includegraphics[width=12cm]{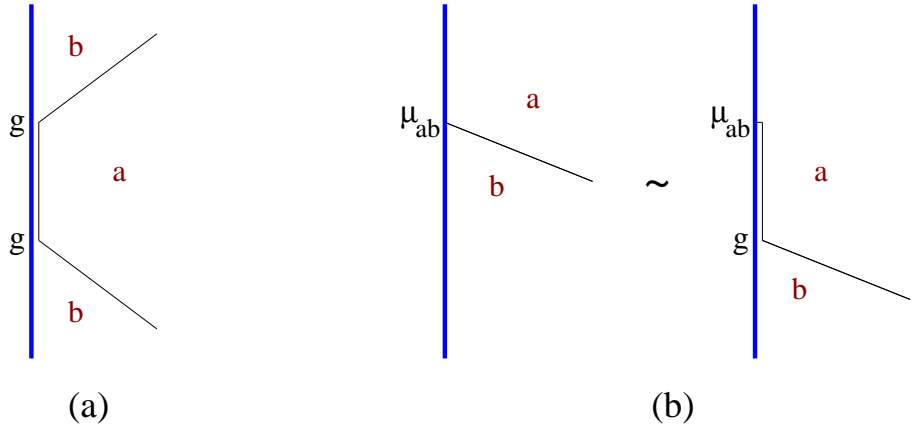}
\caption{The boundary bound state (double line) originating in kink-boundary scattering (a), and a pictorial representation of equation (\ref{ff_pole}) (b).}
\label{wet_bound}
\end{center}
\end{figure}

The boundary bound state affects the results of the previous section for the boundary condition $B_{aba}$ because the leading low-energy contribution in the expansion over intermediate states now comes from $|0\rangle_{B'_{a}}$ rather than from $|K_{ba}(\theta)\rangle_{B_a}$. So the partition function becomes
\EQ
Z={}_{B_a}\langle 0|\mu_{ab}(0,R/2)\mu_{ba}(0,-R/2)|0\rangle_{B_a}=
\left|{}_{B_a}\langle 0|\mu_{ab}(0,0)|0\rangle_{B'_a}\right|^2e^{-mR\cos u}+O(e^{-mR})\,,
\label{Z1}
\EN
and the magnetization profile
\bea
\langle\sigma(x,0)\rangle_{B_{aba}}&\sim &\frac{1}{Z}\,{}_{B_a}\langle 0|\mu_{ab}(0,R/2)|0\rangle_{B'_a}\,{}_{B'_a}\langle 0|\sigma(x,0)|0\rangle_{B'_a}\,{}_{B'_a}\langle 0|\mu_{ba}(0,-R/2)|0\rangle_{B_a}
\nonumber\\
&=&\langle\sigma(x,0)\rangle_{B'_a}\,.
\label{profile3}
\eea
We see then that, as a consequence of (\ref{bulk_magnetization}), the magnetization profile now tends to $\langle\sigma\rangle_{a}$ at large $mx$, in contrast to what obtained in the previous section, where it tended to $\langle\sigma\rangle_b$ for $R$ large enough. This corresponds to the fact that now the asymptotic behavior is determined by the state in which the interface, and then the phase $b$, are bound to the boundary, while before the dominant state was that in which phase $b$ extended to an average midpoint distance of order $\sqrt{R}$ from the boundary.

\begin{figure}[t]
\begin{center}
\includegraphics[width=2.5cm]{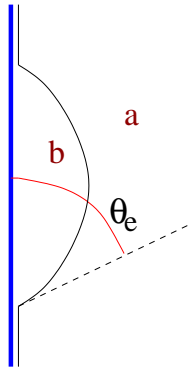}
\caption{Splitting and recombination of the boundary bound state $B_a'$ corresponds to ``partial wetting'', in which a drop of phase $b$ makes an equilibrium contact angle $\theta_e$ with the boundary. Equation (\ref{bs}) with $u=\theta_e$ gives the surface tension balance condition at the contact points.}
\label{wet_contact}
\end{center}
\end{figure}

Consistency of the asymptotic expansion requires that the corrections to (\ref{profile3}) vanish as $R\to\infty$. For $mx$ large, the first of these corrections is that due to the $|K_{ba}(\theta)\rangle$ intermediate states given in (\ref{profile1}). The $Z$ in the denominator, however, is now (\ref{Z1}) rather than (\ref{Z}), so that the correction behaves as $e^{mR(\cos u-1)}$ at large $R$. Hence, if $u$ approaches $0$, i.e. if the interface approaches the unbinding point, consistency requires that $R$ diverges faster than $1/u^2$. If we adopt a vocabulary within which $b$ is a liquid phase and $a$ a vapor phase, we can say that as $u\to 0$ a thin layer of the liquid phase spreads all over the boundary. 

The relation with the usual characterization of interfacial phenomena at boundaries becomes more transparent if we consider the situation usually referred to as ``partial wetting'', corresponding to a drop of liquid sorrounded by a thin layer of liquid adsorbed on the rest of the boundary (see e.g. \cite{BEIMR}). In our formalism this amounts to splitting and recombination of the boundary bound state $B_a'$ (Fig.~3). Considering that the kink mass $m$ is the surface tension of the interface \cite{DV}, that $E_B$ is the surface tension between the boundary and the drop, and that $E_{B'}$ is the surface tension between the boundary away from the drop and phase $a$, we recognize in (\ref{bs}) the Young equilibrium condition at contact points (see e.g. \cite{deGennes} and references therein), with $u$ playing the role of the equilibrium contact angle $\theta_e$ (Fig.~3). In addition, the combination $m(\cos u-1)$ encountered a moment ago is recognized as the so called ``equilibrium spreading coefficient'' (see \cite{BEIMR}). We also see that interface unbinding at $u=0$ corresponds to vanishing of the contact angle, namely to the usual characterization of the wetting transition point (passage from partial to complete wetting).

The boundary bound state is a property of the theory with translationally invariant boundary condition $B_b$. Parameters of this theory are the temperature, related to the kink mass as $m\propto(T_c-T)^\nu$, and a coupling $\lambda$ entering the boundary term $\lambda\int dy\,\phi(0,y)$ of the classical reduced Hamiltonian. If $X$ is the scaling dimension\footnote{The exponents $\nu$ and $X$ are known exactly from bulk \cite{BPZ} and boundary \cite{Cardy} conformal field theory, respectively.} of the boundary field $\phi(0,y)$, $u$ is function of the dimensionless combination $\lambda/m^{1-X}$. If $\lambda$ is kept fixed, the condition $u=0$ determines a wetting transition temperature $T_w(\lambda)<T_c$. 

The results (\ref{Z}), (\ref{Z1}) and (\ref{profile3}) account for those reported in \cite{Abraham_wetting,Abraham} for the particular case of an Ising model with boundary condition $B_{+-+}$ and coupling between the boundary spins and their nearest neighbors different from the coupling within the rest of the lattice; this modified coupling corresponds to the boundary parameter $\lambda$ in this case. The generality of our results also explains why approximated treatments of other models resulted in findings similar to the Ising ones (see \cite{Abraham} and references therein).

\section{Conclusion}
In this paper we studied the scaling limit of a generic ferromagnetic system with a continuous phase transition, below criticality and on the half plane, with boundary conditions favoring one of the phases along an interval of length $R$, and a different phase outside this interval. We used field theory to determine exact large $R$ asymptotics of the magnetization profile perperdicularly to the boundary at the middle of the interval. We showed that, generically, the large $R$ asymptotic behavior corresponds to the presence of an interface pinned at the boundary condition changing points, with an average midpoint distance from the boundary which grows as $\sqrt{R}$. The passage probability density of the interface has the gaussian form found in \cite{DV} for the whole plane, modified by a quadratic factor which accounts for the presence of the boundary. These results are modified if the scattering on the boundary admits a stable bound state, which then becomes leading at low energies and corresponds to the binding of the interface to the boundary. In this case we showed how field theory accounts at a fundamental level for the contact angle and spreading coefficient of the phenomenological wetting theory.

These results follow from general low energy properties of two-dimensional field theory. In particular, the annihilation singularity of the spin field matrix element on one-kink states and the boundary-kink bound state pole play a key role in determining the asymptotics of the magnetization profile in the unbound and bound regimes, respectively.

Additional interfacial properties, such as the internal structure arising from subleading terms of the large $R$ expansion or double interfaces appearing in some models for particular choices of boundary conditions, can be analyzed in the same way it was done in \cite{DV} on the whole plane; we refer the reader to that paper on these points.




\begin{thebibliography}{99}
\bibitem{deGennes} P.G. De Gennes, Rev. Mod. Phys. 57 (1985) 827.
\bibitem{Diehl} H.W. Diehl, in Phase Transitions and Critical Phenomena, edited by C. Domb and J.L. Lebowitz,  Vol. 10, p.~75, Academic Press, London, 1986. 
\bibitem{Dietrich} S. Dietrich, in Phase Transitions and Critical Phenomena, edited by C. Domb and J.L. Lebowitz,  Vol. 12, p.~1, Academic Press, London, 1988.
\bibitem{Schick} M. Schick, in Liquids at Interfaces, edited by J. Chavrolin, J.-F. Joanny and J. Zinn-Justin, p. 415, Elsevier, Amsterdam, 1990.
\bibitem{FLN} G. Forgacs, R. Lipowsky and T.M. Nieuwenhuizen, in Phase Transitions and Critical Phenomena, edited by C. Domb and J.L. Lebowitz,  Vol. 14, Chap.~2, Academic Press, London, 1991.
\bibitem{BR} D. Bonn and D. Ross, Rep. Prog. Phys. 64 (2001) 1085.
\bibitem{BLM} K. Binder, D.P. Landau and M. Muller, J. Stat. Phys. 110 (2003) 1411.
\bibitem{BEIMR} D. Bonn, J. Eggers, J. Indekeu, J. Meunier and E. Rolley, Rev. Mod. Phys. 81 (2009) 739.
\bibitem{Abraham_wetting} D.B. Abraham, Phys. Rev. Lett. 44 (1980) 1165.
\bibitem{Abraham} D.B. Abraham, in Phase Transitions and Critical Phenomena, edited by C. Domb and J.L. Lebowitz,  Vol. 10, p.~1, Academic Press, London, 1986.
\bibitem{DV} G. Delfino and J. Viti, J. Stat. Mech. (2012) P10009.
\bibitem{ELOP} R.J. Eden, P.V. Landshoff, D.I. Olive, J.C. Polkinghorne, The Analitic S-Matrix, Cambridge University Press, 1966.
\bibitem{GZ} S. Ghoshal and A.B. Zamolodchikov, Int. J. Mod. Phys. A9 (1994) 3841; Erratum, ibidem A9 (1994) 4353.
\bibitem{BPT} Z. Bajnok, L. Palla and G. Takacs, Nucl. Phys. B 750 (2006) 179.
\bibitem{Smirnov} F.A. Smirnov, Form Factors in Completely Integrable Models of Quantum Field Theory, World Scientific, 1992.
\bibitem{DC98} G. Delfino and J. Cardy, Nucl. Phys. B  519 (1998) 551.
\bibitem{AI} D.B. Abraham and M.E. Issigoni, J. Phys. A 13 (1980) L89.
\bibitem{BPZ} A.A. Belavin, A.M. Polyakov, A.B. Zamolodchikov, Nucl. Phys. B 241(1984) 333.
\bibitem{Cardy} J. Cardy, Nucl. Phys. B 240 (1984) 514.

\end{thebibliography}
\end{document}